\documentclass[referee]{raa}            % referee version: for submission

\usepackage{graphicx,times}             %for PS/EPS graphics inclusion, new
\usepackage{enumerate}

\newcommand{\aspc}{   {\it Astron. Soc. Pac. Conf. Series} }

\begin{document}

\title{On the line profile changes observed during the X2.2 class flare in the active region NOAA 11158}

%   \subtitle{I. Place Your Subtitle Here}

   \volnopage{Vol.0 (200x) No.0, 000--000}      %%preserved for Editor. DOn't remove!
   \setcounter{page}{1}          %%starting page, preserved for Editor. DOn't remove!

   \author{A. Raja Bayanna
      \inst{1}
   \and Brajesh Kumar
      \inst{1}
   \and P. Venkatakrishnan
      \inst{1}
   \and Shibu K Mathew
      \inst{1}
   \and B. Ravindra
      \inst{2}
   \and Savita Mathur
      \inst{3}
   \and R. A. Garc\'ia
      \inst{4}
   }

   \institute{Udaipur Solar Observatory, Physical Research Laboratory, Dewali, Badi Road,
Udaipur 313 004, India {\it bayanna@prl.res.in}\\
        \and
             Indian Institute of Astrophysics, II Block, Koramangla, Bangalore 560034, India\\
        \and
             Space Science Institute, 4750 Walnut street Suite\#205, Boulder, CO 80301, USA\\
	\and
	     Laboratoire AIM, CEA/DSM-CNRS, Universit\'e Paris 7 Diderot, IRFU/SAp, Centre de Saclay, 91191, Gif-sur-Yvette, France
   }

   \date{Received~~2009 month day; accepted~~2009~~month day}

\abstract{ The solar active region NOAA 11158 produced a series of flares during its passage through the solar disk. The first major flare (of class X2.2) of the current solar cycle occurred in this active region on 2011 February 15 around 01:50 UT. We have analyzed the Dopplergrams and magnetograms obtained by the Helioseismic and Magnetic Imager (HMI) instrument onboard {\it Solar Dynamics Observatory} ({\it SDO}) to examine the photospheric velocity and magnetic field changes associated with this flare. The HMI instrument provides high-quality Doppler and magnetic maps of the solar disk at 0.5 arcsec spatial scale at a cadence of 45~s along with imaging spectroscopy. We have identified five locations of velocity transients in the active region during the flare. These transient velocity signals are located in and around the flare ribbons as observed by {\it Hinode} in Ca~II~H wavelength and the footpoints of hard X-ray enhancement in the energy range 12-25 keV from RHESSI. The changes in shape and width of two circular polarization states have been observed at the time of transients in three out of five locations. Forward modeling of the line profiles shows that the change in atmospheric parameters such as magnetic field strength, Doppler velocity and source function could explain the observed changes in the line profiles with respect to the pre-flare condition. 
}

   \authorrunning{A. Raja Bayanna et al.}            %author_head in even pages
   \titlerunning{Line profile changes observed in NOAA 11158}  % title_head in odd pages

   \maketitle
%% The author head (on even pages) and the title head (on odd pages) will be
%% automatically extracted from \author{} and \title{}. Whenever the title is too long,
%% you will be asked to supply a shorter one by inserting either \authorrunning{} or
%% \titlerunning{} before \maketitle. Anyway, you can specify your own heads.
%%
%%
%% Note: In the following text body of your manuscript, please note several differences from
%%       other major journals:
%% (1) \subsection{Please Capitalize the First Letter of Each Notional Word in Subsection Title}
%% (2) Please Capitalize the First Letter of Each Notional Word in all tables' captions

%
%________________________________________________ sections below
%
\section{Introduction}           %% first-level sections will be auto-capitalized
\label{sect:intro}

In solar flares, the magnetic energy stored in the active regions is catastrophically released in the form of large amount of thermal energy and energetic particles moving with relativistic speeds. Investigations have been made to study the flare related changes in various magnetic properties associated with the active regions. Significant changes are known to take place in magnetic shear (Hagyard et al.~\cite{Hagyard1999}; Tiwari et al. \cite{Tiwari2010}), magnetic flux (Sudol \& Harvey~\cite{sudolHarvey2005}; Wang et al.~\cite{Wang2009}; Wang et al.~\cite{Wang2011}; Wang et al.~\cite{Wang2012}), magnetic helicity ( Moon et al.~\cite{Moon2002}; Ravindra et al.~\cite{Ravindra2011}) and magnetic energy (Metcalf et al.~\cite{Metcalf2005}; Ravindra \& Howard~\cite{Ravindra2010}; Sun et al.~\cite{Sun2012}) during the flares.

Apart from these observed changes in the magnetic field properties during the flare, it is also believed that the solar flares can excite oscillations in the Sun. Wolff~(\cite{Wolff1972}) suggested that large solar flares would cause a thermal expansion that would launch a wave of compression to move sub-sonically into the solar interior, thereby stimulating solar oscillations. Following this, several researchers have made an attempt to study the effect of flares on the acoustic velocity oscillations of the Sun discovered in the 1960s (Leighton et al.~\cite{Leighton62}). These global {\it p}-mode oscillations are excited by the turbulence in the convective layers of the Sun (Leibacher \& Stein~\cite{Leibacher71}, Goldreich et al.~\cite{Gold1994}). Haber et al.~(\cite{Haber1988}) reported increase in the average power of intermediate degree modes, after a flare. The first detection of `solar quakes' inside the Sun was reported by Kosovichev \& Zharkova~(\cite{KosovichevZharkova1998}) during the X2.6 class flare on 1996~July~9. Since then, there have been several reports on the local and global effects of flare on the solar velocity oscillations (Kumar et al.~\cite{kumar10}; Kumar et al.~\cite[and references therein]{kumar11}).

In this paper, we present a detailed study of the flare associated velocity and magnetic field changes observed in the active region NOAA 11158 during the X2.2 class flare on 2011 February 15 using the full-disk Dopplergrams and magnetograms obtained by the Helioseismic and Magnetic Imager (HMI; Schou et al.~\cite{HMI2011}) onboard {\it Solar Dynamics Observatory} ({\it SDO}; Pesnell et al.~\cite{SDO2011}) spacecraft. The active region NOAA 11158  was located close to the disk center (S21, W21). The active region comprised of four sunspots, central one having $\delta$-type magnetic configuration which produced a X2.2 class flare on 2011 February 15. The active region also produced a few M-class flares in the subsequent days in addition to few weaker flares prior to this flare-event. As seen in the Geostationary Operational Environmental Satellite ({\it GOES}) soft X-ray flux (1--8~\AA) data, the flare started at 01:44 UT, peaked at 01:56 UT and ended at 02:06 UT. Wang et al.~(\cite{Wang2012}) showed that strong magnetic shear developed in this active region caused by the rapid evolution of the horizontal magnetic field and it played an important role for the trigger of this X-class flare. Vemareddy et~al.~(\cite{Vema2012}) and Song et~al.~(\cite{Song2013}) have studied the magnetic non-potentiality parameters of this active region and concluded that rotational motion of the sunspot played a significant role in increasing the non-potentiality thereby leading to the observed multiple flares and CME event in this active region.

Using {\it SDO}/HMI data, Martinez et al.~(\cite{Couvadit2011}) carried out line profile analysis of a white-light M-class flare for which they observed a step-wise change in the line-of-sight magnetic field at the flare footpoints. They conclude that the line remains in absorption with a blue shift at both the footpoints without any change in the line width and shape. In this event of X2.2-class flare, we have found five different locations in the active region which show enhanced velocity signals during the flare. These locations also show significant changes in magnetic signals during the flare. On contrary to the observations of Martinez et al.~(\cite{Couvadit2011}) for the M-class flare, here we observe that three out of the five locations show line profile changes. We use forward modeling of the pre-flare profiles of the two circular polarization states ({\it LCP} and {\it RCP}) to show that the observed changes in the line profiles during the flare can indeed be interpreted principally due to magnetic and velocity changes. 

\section{The observational data}

In this work, we use photospheric observations of the active region NOAA 11158 centered at 6173~\AA . The data consists of the full-disk images of intensity, magnetic field and velocity at a cadence of 45~s for the period starting from 01:00 UT to 02:30 UT obtained from the HMI instrument onboard {\it SDO}. We have also used {\it LCP} and {\it RCP} data from HMI for examining the line profile changes during the flare. HMI observes both these polarizations at six wavelength positions along the line profile, sequentially, at an approximate cadence of 3.75~s giving an overall cadence of 45~s.

From Solar Optical Telescope (SOT; Tsuneta et al.~\cite{tsun08}) onboard {\it Hinode} (Kosugi et al.~\cite{kosu07}), we have obtained filtergrams of Ca~II~H centered at 3968~\AA, recorded at a cadence of 5 min starting from 01:00 to 02:30 UT. However, the observing cadence of the instrument during the period from 01:50 to 02:02 UT was set as 20~s to have a better observation of the temporal evolution of the flare. The spatial scale of these Ca~II~H images is $\sim$0.1 arcsec per pixel. These filtergrams reveal that the flare is a typical two-ribbon type and the ribbons lie on both sides of the polarity inversion line. These observations are used to identify the location of flare kernels.

We have also used the hard X-ray (HXR) spectral data from {\it Reuven Ramaty High Energy Solar Spectroscopic Imager (RHESSI;} Lin~et~al.~\cite{Lin2002}{\it)}  in the energy band 12-25~keV to compare the temporal evolution of hard X-ray peak with the spatial coincidence of the velocity and magnetic field changes during the flare.

\begin{figure}[!t]
\centering
\includegraphics[width=0.8\textwidth, angle=0]{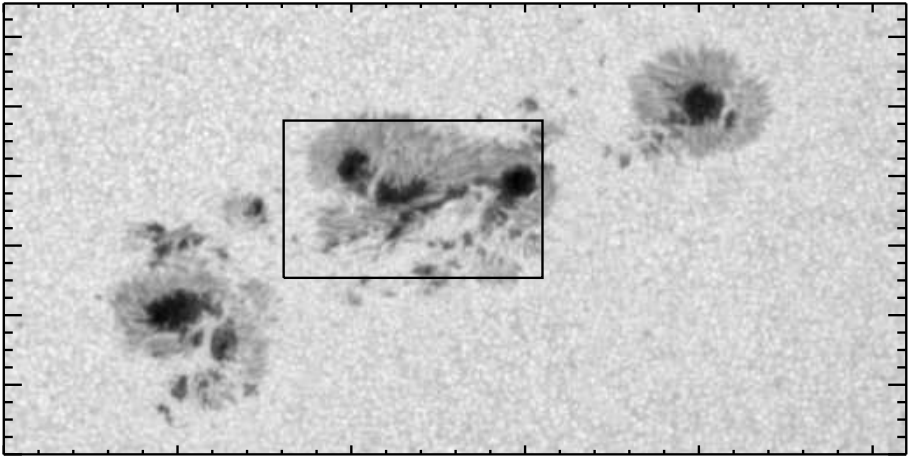}
\caption{The continuum image of the active region NOAA 11158 as observed with {\it SDO}/HMI. The box shown in thick lines contains the region which produced the X2.2 class flare. The total field-of-view is 260$\times$130~arcsec$^2$.}
\end{figure}

\section{Data analysis}

A grid of area ~260$\times$130 arcsec$^2$ containing the active region NOAA 11158 (c.f. Figure~1) is extracted from the full-disk intensity image of HMI. The raster of these intensity images were co-aligned using a two-dimensional cross-correlation technique to a sub-pixel accuracy. Later on, the co-aligned raster of these intensity images were used to co-align the co-temporal Dopplergrams, magnetograms, and images representing the circular polarizations from HMI. The Ca~II~H filtergrams obtained from {\it Hinode}/SOT are also co-aligned with respect to intensity images obtained from HMI by reducing the pixel resolution from 0.1 arcsec to 0.5 arcsec.

The RHESSI images are reconstructed using the CLEAN-algorithm (Hurford et~al.~\cite{clean2002}) for 1.0~arcsec per pixel spatial scale and further  interpolated to 0.5 arcsec per pixel to match with the HMI image scale. The FOV of the reconstructed images is 128 arcsec $\times$ 128 arcsec. These images are co-aligned with the HMI data sets using the header information available in both the data. However, because of low-resolution nature of the RHESSI images, the accuracy of the co-alignment of RHESSI images with respect to the images obtained by HMI is about 1.0~arcsec.

\subsection{Identification of localized velocity transients}
A two-point backward difference filter (Garc\'ia \& Ballot~\cite{garc08}) is applied to the time series of images to remove the slow varying solar background features. Thus, we obtain difference image series of the Dopplergrams. Using this difference image sequence, we construct a `depth image' by obtaining the difference between the minimum and maximum values of the time series for each pixel. From this, we identified the pixels with the velocity values larger than 1~km/s as the locations of large velocity flows during the flare.

We have also constructed two root-mean-square (rms) maps. One is from a sequence of Dopplergrams obtained before the flare starting from 01:10 to 01:40 UT. Another one is from Dopplergrams obtained during the flare event starting from 01:40 to 02:10 UT. A difference image was constructed from these two rms images. The pixels with enhanced velocity in the active region were identified by using a threshold velocity of $\pm$ 200 m/s in the aforesaid difference of rms images of the Dopplergrams. The threshold of $\pm$200~m/s is equivalent to $\pm$3$\sigma$ that of the quiet Sun.

\begin{figure}[!t]
\centering
\includegraphics[width=0.7\textwidth]{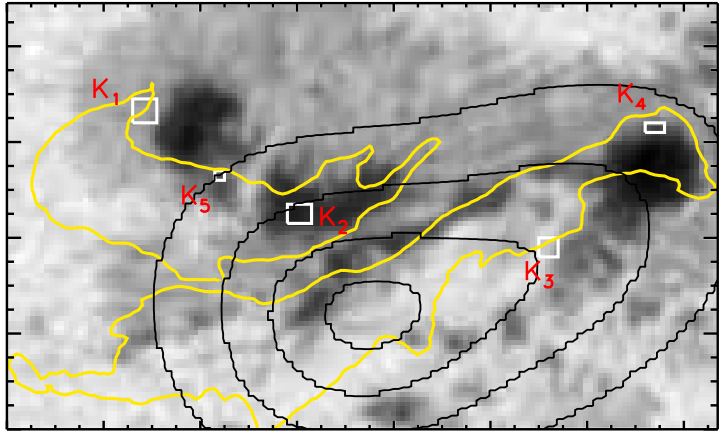}
\caption{The continuum image showing the part of the active region NOAA 11158, which is used in our analysis. $K_1$, $K_2$, $K_3$, $K_4$, and $K_5$ represent the kernels of the significantly enhanced velocity signals. The yellow contours show the location of flare ribbons extracted from the temporally averaged Ca~II~H images spanning the period 01:50--02:00~UT. The total field-of-view is ~75$\times$45 arcsec$^2$. Contours of temporally averaged HXR image from {\it RHESSI} spanning the period 01:50--02:00~UT are shown in black solid lines. The contours are corresponding to the 95\%, 75\%, 50\%, and 25\% of the maximum value, respectively.}
\end{figure}

Using the above criteria, we selected a set of pixels showing enhanced velocity signals. All of these are located in the central sunspot of the active region, which produced the X2.2 class flare on 2011 February 15. We grouped all these pixels into five regions (or, kernels), namely, $K_1, K_2, K_3, K_4$, and $K_5$ as shown in Figure~2. Here, the kernel $K_5$ contains the weakest velocity transient relative to $K_1, K_2, K_3$, and $K_4$. For comparison, we considered a quiet region of 3$\times$3 pixels away from the active region. The regions of velocity transients $K_1$, and $K_2$ identified by us coincide with the locations of seismic sources shown by Kosovichev~(\cite{kosovichev2011}), $K_4$ coincides with the seismic source shown by Zharkov et al.~(\cite{Zharkov13}). In addition to this the location $K_2$ coincides with the transient velocity location shown by Maurya et al.~(\cite{Maurya2012}) as well. 

As indicated earlier, a raster of the active region along with the locations of kernels with transient velocity signals are shown in  Figure~2. The RHESSI HXR contours corresponding to the 95\%, 75\%, 50\%, and 25\% of the maximum value of the temporally averaged image during the period 01:50--02:00~UT are overlaid on the intensity image. The time period 01:50--02:00~UT is considered for constructing the average image as this duration spans the impulsive phase of the flare as seen in the HXR light curve from {\it RHESSI}. The velocity and magnetic changes are also observed during this period. Here, we report that the kernels $K_2$, $K_3$, $K_4$ and $K_5$ spatially coincide with the location of HXR enhancement. It is important to note that all these kernels are located in the weaker HXR emission areas, while the kernel $K_1$ is slightly away from location of enhancement in HXR. In this regard, it is to be noted that Kosovichev~(\cite{kosovichev2011}) has already pointed out that the kernel $K_1$ (or Impact 1) appears before the impulsive phase of the HXR emission. 

We also overlaid the contours of the flare ribbons seen in the Ca~II~H image constructed as a temporal average over the duration 01:50--02:00~UT on the intensity image of the active region as observed by {\it SDO}/HMI. All the kernels, $K_1, K_2, K_3, K_4$, and $K_5$ are located in and around the flare-ribbons as evident from Figure~2. However, it is interesting to note that these transient signals are not located in the central part of the flare ribbons.

In order to examine the behavior of the velocity signals within the flare ribbons with respect to a non-flaring magnetic region, we plot a histogram (c.f., Figure 3) of rms Doppler velocity of all the pixels contained in the flare ribbons for the epochs before, during and after the flare. We observe that the rms value of the Doppler velocity for the aforementioned pixels is higher during the flare in comparison to that before and after the flare. A similar analysis is done for another sunspot group (right most in Figure~1) in the same active region and also in a quiet region. Both show no change in the histograms during the flare compared to before and after the flare.

\begin{figure}[!h]
\centering
\includegraphics[width=0.36\textwidth, angle=90]{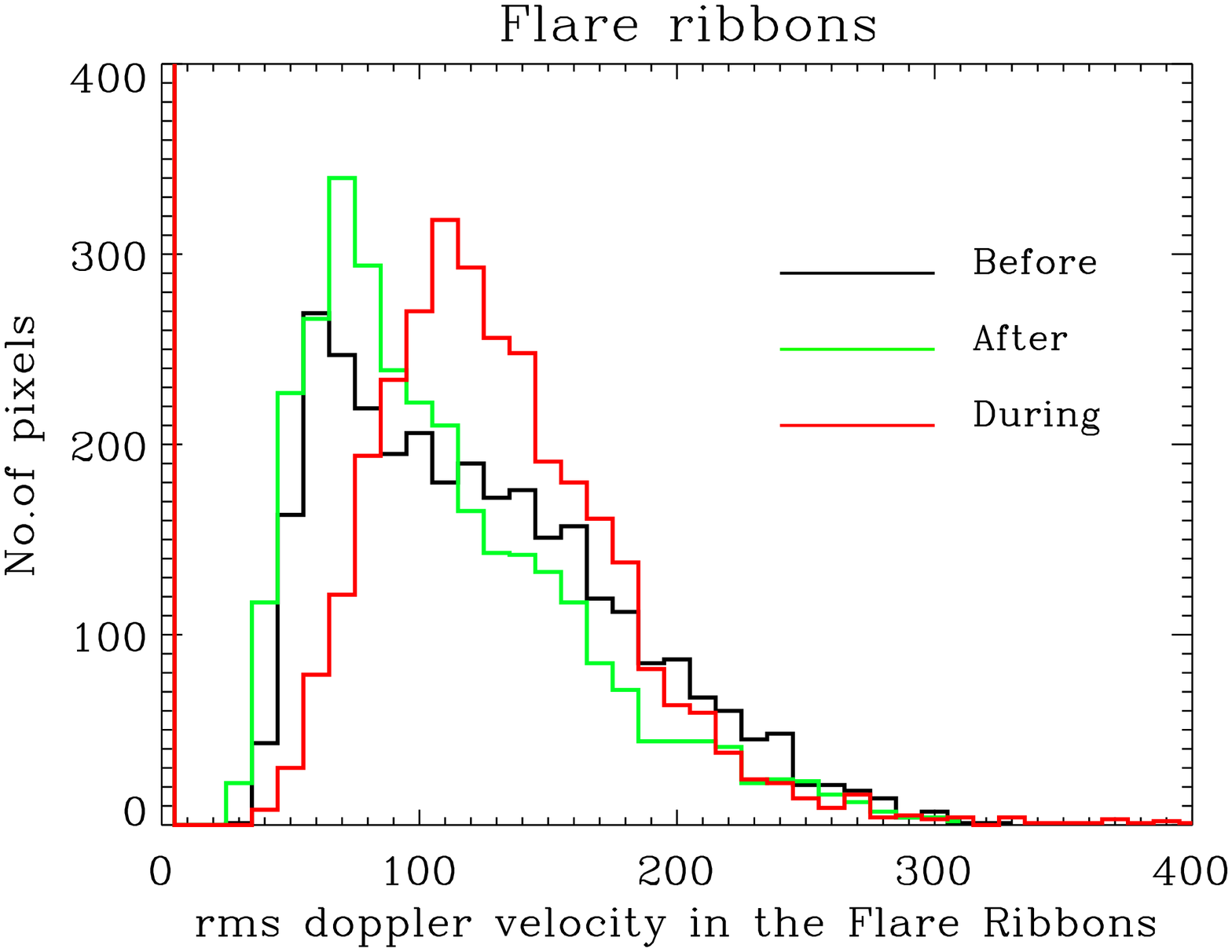}
\includegraphics[width=0.36\textwidth, angle=90]{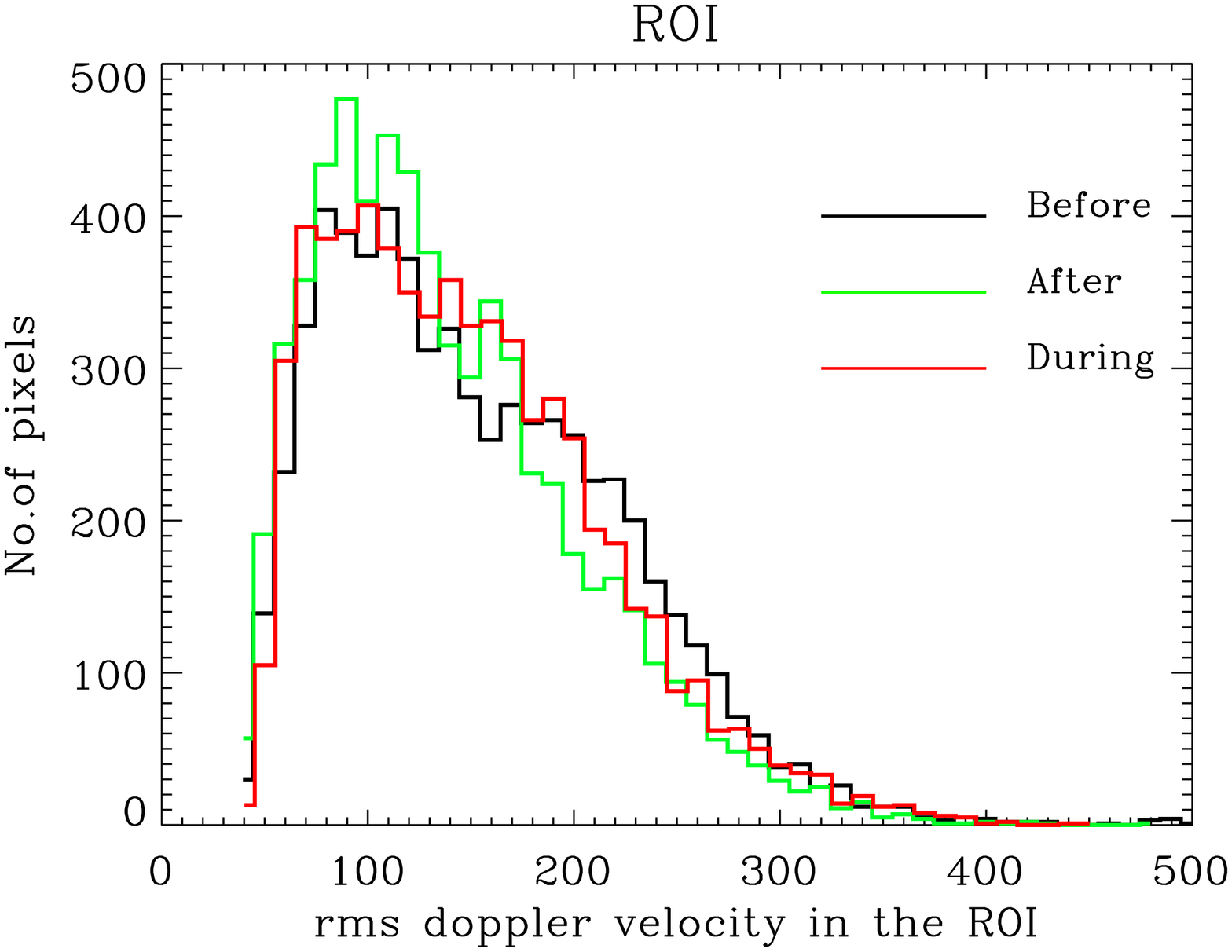}
\caption{Left: Histogram of rms Doppler velocity of all the pixels contained in the flare ribbons for the epochs before (black), during (red) and after (green) the flare. Right: Histogram of rms Doppler velocity for non-flaring sunspot group (right most in Figure~1) in the same active region for the epochs ~before (black), during (red) and after (green) the flare.}
\end{figure}

\subsection{Temporal variations of velocity and magnetic fields in the identified locations of velocity transients}

\begin{figure}[!t]
\begin{center}
\vspace{-15mm}
\includegraphics[width=14.5cm]{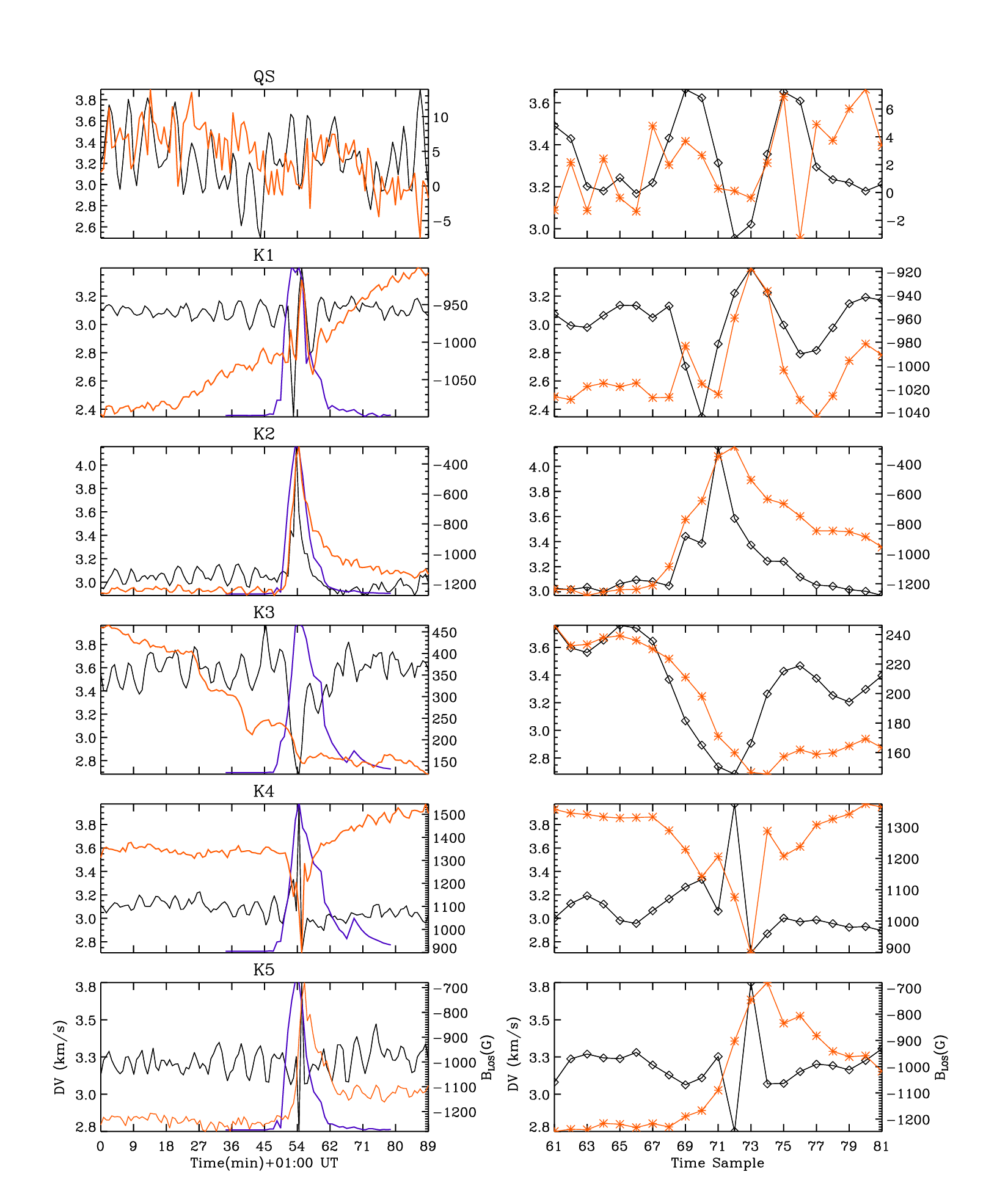}
\caption{Left: Temporal evolution of Doppler velocity (DV) and $B_{LOS}$ (over-plotted orange color) for the duration 01:00-02:30 UT. Right: the same data for the period 01:45-02:00 UT  of the kernels $K_1, K_2, K_3, K_4$, and $K_5$. The over-plotted blue curve shows the temporal evolution of hard X-ray flux from {\it RHESSI} in the energy band 12-25~keV corresponding to the localized positions. The panels in the top show the DV and $B_{LOS}$ for a quiet area away from the active region.}
\label{smoothvar}
\end{center}
\end{figure}

In Figure~4, we show the time series of Doppler velocity (DV) and magnetic field ($B_{LOS}$) variations over five locations ($K_1, K_2, K_3, K_4$, and $K_5$) and one of the quiet Sun locations. As evident from Figure~4, almost all the regions show sudden jump in the DV as well in the $B_{LOS}$ during the flare, except $K_3$, where we observe a gradual decrease in the $B_{LOS}$ from $450$~G to $120$~G. In general, DV returns back to the pre-flare levels after the flare but a permanent change of approximately 100~G is observed in the case of $B_{LOS}$. Such behavior of step-function-like changes (or, rapid changes) and gradual changes in magnetic fields during M-class and X-class flares have been reported earlier by several researchers (Kosovichev \& Zharkova~\cite{KosoZhar99};  Kosovichev \& Zharkova~\cite{KosoZhar01}; Sudol \& Harvey~\cite{sudolHarvey2005}; Wang~\cite{Wang2007}; Wang et al.~\cite{Wang2009}; Zhao et al.~\cite{Zhao09}; Wang et al.~\cite{Wang2011}). In our analysis, it is also observed that in all the cases velocity transients occur prior to the magnetic transients (c.f., Table 1) and the time of velocity and magnetic transients are separated at least by 45~s.
\begin{table}[t]
\begin{center}
\caption{\bf Time (UT) of velocity and magnetic transients in the identified kernels}
\vspace{5mm}
\begin{tabular}{|c|c|c|}
\hline
Kernel&   Velocity   & Magnetic field (LOS)\\
\hline
$K_1$&01:52:24& 01:54:39\\
$K_2$&01:53:09& 01:53:54\\
$K_3$&01:53:54& 01:55:24\\
$K_4$&01:53:54& 01:54:39\\
$K_5$&01:53:54& 01:55:24\\
\hline
\end{tabular}
\end{center}
\end{table}

In order to examine the evolution of hard X-ray light curve during the flare corresponding to the localized positions $K_1, K_2, K_3, K_4$, and $K_5$, we have over-plotted the {\it RHESSI} spectral data (blue-dotted curves) in the energy range 12-25~keV in Figure~4. It is observed that the hard X-ray energy peaks either at 01:53~UT or 01:54~UT in all these locations. For the  kernels, $K_2, K_3, K_4$, and $K_5$, we observe that the peaks of velocity transients are seen within a minute from the peak time of HXR evolution at these locations in the energy band 12-25~keV. This result is compatible with the theoretical simulations of Zharkova \& Zharkov (\cite{ZharZhar07}). However, the peak time of velocity transients observed at the kernel $K_1$ appears during the peak time of the hard X-ray variations. It is to be noted that the HXR light-curve in the energy band 12-25~keV at the kernel $K_1$ has a structured peak. Hence, it is difficult to conclude in this case whether HXR impulse precedes the velocity transients.

\subsection{Line profile changes associated with the identified locations of velocity and magnetic transients}

It is understandable that the large velocity and strong magnetic field can cause line profile of Fe~I to be out of working range of the {\it SDO}/HMI inversion algorithm, which is used for deriving the line-of-sight (LOS) velocity and magnetic field measurements. In this regard, limitation of HMI algorithm depends on the surface velocity, magnetic field strength, orbital velocity of the space craft and the location of the region of interest on the solar disk. As the total wavelength range covered by HMI is only about 344~m\AA, the Fe~I line moves out of working range of HMI algorithm for Doppler velocities larger than $\pm$8.3 km/s or an equivalent magnetic field strength of 3800 G, which yields bad result (Couvidat et al.~\cite{Couvidat2012}). In all the kernels identified by us, the maximum velocity including orbital velocity of the spacecraft is 4.2~km/s and maximum field strength is less than 1500~G. Thus, the observed LOS velocity and magnetic field strength are within the working range of the instrument. 

On the other hand, in the absence of spectro-polarimetric observations, it has been always a concern that large velocity flows or rapid changes in magnetic fields observed during the flares could be a result of artifacts arising due to distortions in the line profiles (Ding et al.~\cite{Ding2002}, Qiu \& Gary~\cite[and references therein]{QaG2003}). Wang et al.~(\cite{Wang2011}), for the first time, analyzed spectral line profiles of a flaring region which showed abrupt magnetic flux changes during three successive M-class solar flares in the active regions NOAA 10039 and NOAA 10044 observed on 2002 July 6. They used data observed by the Imaging Vector Magnetograph (IVM) at Mees solar observatory to check possible line profile changes during the flares. They observed that the fluctuations in the width, depth and central wavelength of the lines were less than 5\% and also they ruled out any obvious signature of line profile changes associated with the flare. Similarly, Martinez et al.~(\cite{Couvadit2011}) did not find any change in the line width and shape during a white-light M-class flare using the spectro-polarimetric data from {\it SDO}/HMI. Zhao et al.~(\cite{Zhao09}) attempted to quantify the flare-induced signals during an X2.6 class flare observed on 2005 January 15, and studied the correlation of the flare-induced signal with real magnetic changes during the flare. They used the BBSO vector magnetic field data along with co-temporal Ca~I~6103~\AA~and~$H_{\alpha}$ observations for their study. They have found that the flare-induced signal appear in the form of apparent polarity reversal in the observed magnetograms. Their analysis also revealed that the flare-induced signal appeared both in circular polarization measurements and transverse magnetograms.

Here, we analyze the spectro-polarimetric profiles for this X2.2 class flare obtained from {\it SDO}/HMI corresponding to the times of the velocity and magnetic transients at the identified locations. HMI observes the {\it LCP} and {\it RCP} at six wavelength positions in the wavelength range 6173~\AA$~\pm~$172~m\AA, sequentially, at a cadence of 45 s. These polarization data were reproduced by HMI team by interpolating the raw data to a target time as required.

In order to understand the flare associated changes in the line profiles of the identified kernels, we examined co-temporal {\it LCP} and {\it RCP} profiles for this flare similar to Martinez et al.~(\cite{Couvadit2011}) for an M-class flare. We used Gaussian fit with four parameters (continuum intensity ($I_c$), line depth ($I_d$), line center wavelength ($\lambda_0$), and  FWHM ($\sigma$); c.f. equation 3 of Couvidat et al.~\cite{Couvidat2012}) for fitting the six measurements ($I_i,~i=0,5$) of  {\it LCP} and {\it RCP}. Initial values of $I_d$ and $\sigma$ are obtained from Fourier coefficients (c.f. equations 6 and 7 of Couvidat et al.~\cite{Couvidat2012}). The initial value of $I_c$ is obtained from the maximum of the $I_i$'s, and the wavelength position where the minimum of $I_i$'s occur is chosen as the initial value of $\lambda_0$. This results in an optimum Gaussian fit to the observed {\it LCP} and {\it RCP} measurements with minimized chi-square value. 

\begin{figure}[!h]
\begin{center}
\includegraphics[width=3cm]{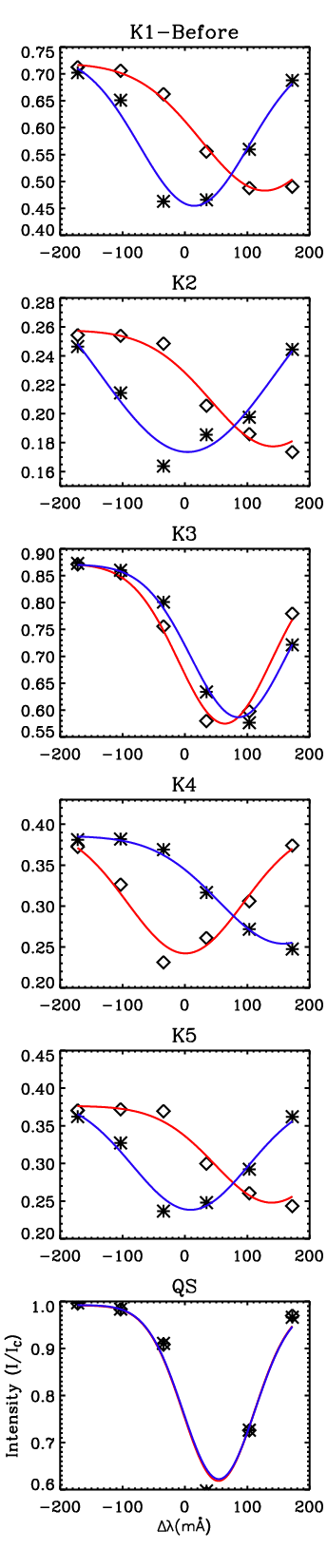}
\hspace{1mm}
\includegraphics[width=3cm]{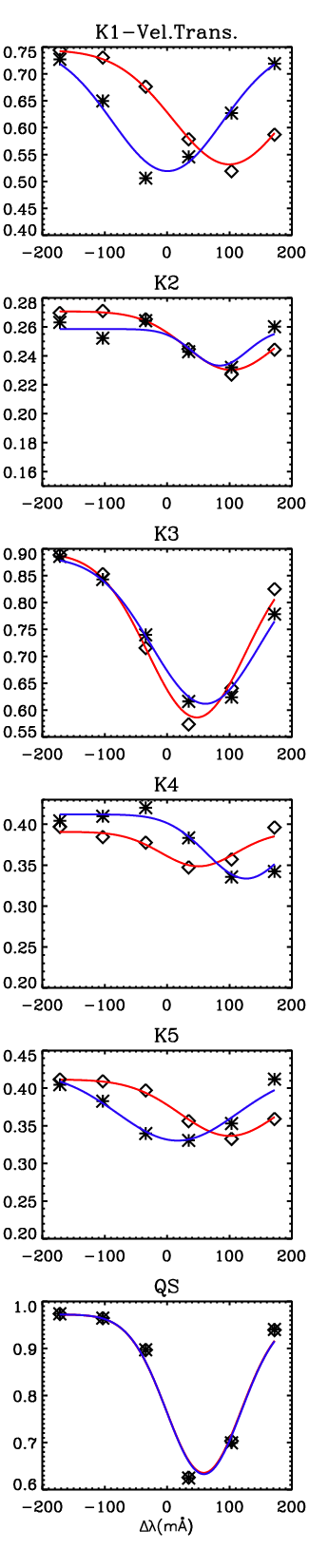}
\hspace{1mm}
\includegraphics[width=3cm]{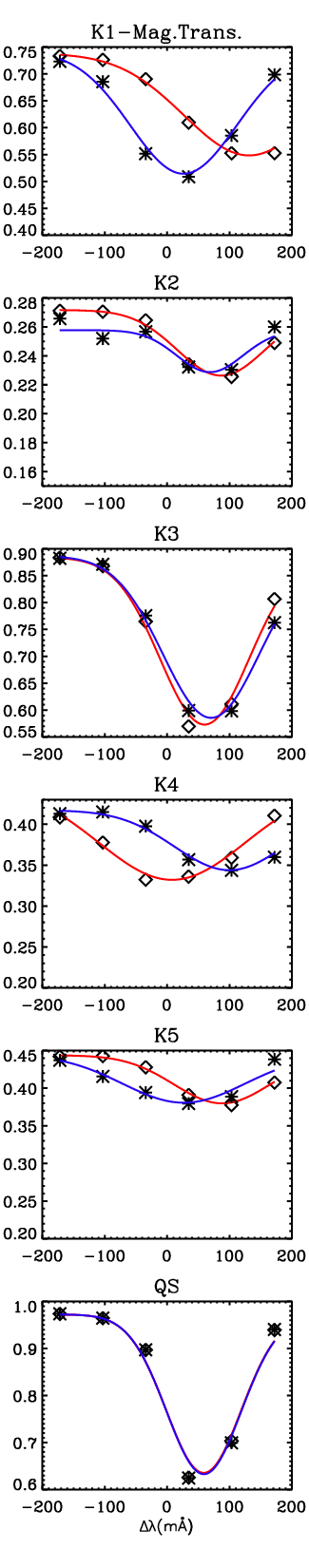}
\hspace{1mm}
\includegraphics[width=3cm]{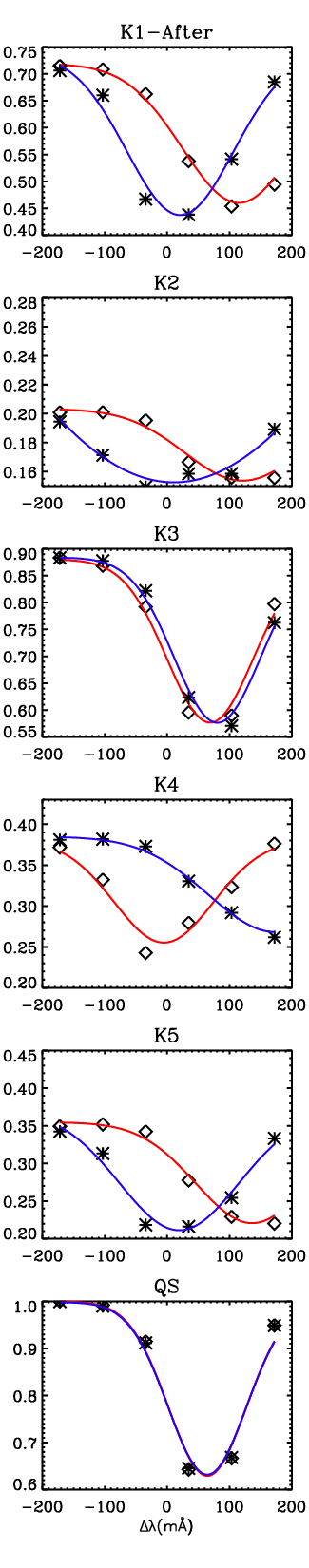}
\caption{{\it LCP} (blue) and {\it RCP} (red) profiles of Fe I line for the epochs: pre-flare (first column), time of velocity transient (second column), time of magnetic transient (third column) and post-flare (fourth column) for the kernels $K_1$, $K_2$, $K_3$, $K_4$, and $K_5$ and quiet Sun (QS), respectively, from top to bottom. Symbols show observed spectra and the continuous lines show the corresponding spectral fit.}
\label{trans}
\end{center}
\end{figure}

Based on the above procedure to fit the six measurements, we now examine the line profile changes in all the kernels. Figure~\ref{trans} shows the {\it LCP} (star symbol) and {\it RCP} (diamond symbol) profiles along with their respective spectral fits (blue and red curves) for the epochs: before (column~I), at the time of velocity (column~II) and magnetic (column~III) transients and after the flare (column~IV). In Table~2, we report the line depth, full-width at half maximum, line center of the profiles and chi-square value of the fits for all the epochs. From Figure~\ref{trans}, it is evident that there are no significant changes in the shape and width of the line profiles at the time of transients for the kernels $K_1$, and $K_3$ . The kernels $K_2$, $K_4$ and, $K_5$ show a change in shape (depth and width) of line profile as well as local increase in intensity at $-$34~m\AA~ and $+$172~m\AA. These wavelength positions are on the wings of the line profile (near the inflection points of the line profiles).

\begin{table}[!h]
\begin{center}
\caption{\bf Parameters of {\it LCP} and {\it RCP} of the kernels at the time of velocity and magnetic transients}
\vspace{5mm}
\begin{tabular}{|c|l|r|r|r|r|r|r|}
\cline{1-8}
Kernel&Flare    & \multicolumn{2}{|c|}{Line Center}& \multicolumn{2}{|c|}{Line Width} &\multicolumn{2}{|c|}{${\chi}^2$ $\times$$10^{-4}$}\\
      &time     & \multicolumn{2}{|c|}{(m\AA)}     & \multicolumn{2}{|c|}{(m\AA)} &\multicolumn{2}{|c|}{} \\
\cline{3-8}
      &         &  {\it LCP} & {\it RCP} & {\it LCP} & {\it RCP} & {\it LCP} & {\it RCP}  \\
\cline{2-8}
$K_1$ &Before   &       14.6&      128.1&      191.8&      239.1&      1.36&     0.18\\
&Vel. Transient & 0.8  & 099.7& 186.6& 213.3 & 1.40 & 0.10\\
&Mag. Transient &       25.8&      131.6&      191.8&      244.2&     0.32&    0.08\\
      &After    &       20.6&      114.4&      191.8&      211.6&      1.04&     0.16\\
\cline{2-8}
$K_2$ &Before   &      4.3&      141.0&      225.3&      233.9&     0.28&     0.11\\
&Vel. Transient &       84.3 & 102.3& 104.9& 170.3 & 0.08 & 0.01\\
&Mag. Transient &       68.8&      86.9&      124.7&      166.8&    0.08&    0.01\\
      &After    &       11.2&      121.3&      255.4&      220.2&    0.07&    0.04\\
\cline{2-8}
$K_3$ &Before   &       86.0&      63.6&      180.6&      175.4&     0.30&     0.40\\
&Vel. Transient & 61.0 & 47.3 & 200.4& 183.2 & 0.32 & 0.42\\ 
      & Mag. Transient&       70.5&      60.2&      182.3&      168.6&     0.35&     0.53\\
      &After    &       80.8&      68.8&      164.2&      163.4&     0.55&     0.62\\
\cline{2-8}
$K_4$ &Before   &       157.4&     0.86&      242.5&      192.6&    0.08&     0.55\\
&Vel. Transient & 126.4& 50.7 & 149.6& 142.7 & 0.19 & 0.11\\
      &Mag. Transient&       99.8&      7.7&      206.4&      216.7&    0.05&    0.08\\
      &After    &       172.0&     -5.2&      251.1&      177.2&    0.060&     0.53\\
\cline{2-8}
$K_5$ &Before   &       8.6&      139.3&      196.9&      215.0&     0.34&     0.31\\
&Vel. Transient& 16.3 & 100.6& 203.8& 187.5 & 0.15 & 0.02\\
      &Mag. Transient&       26.7&      87.7&      211.6&      180.6&     0.18&    0.020\\
      &After    &       18.9&      135.9&      203.8&      211.6&     0.36&     0.16\\
\cline{2-8}
$QS $ &Before   &       54.2&      54.2&      136.7&      135.8&      2.21&      2.23\\
      &Transient&       58.5&      58.5&      141.9&      141.9&      1.52&      1.50\\
      &After    &       64.5&      64.5&      148.8&      147.9&      1.47&      1.50\\
\hline
\end{tabular}
\end{center}
\end{table}

In what follows, we demonstrate, using forward modeling that these changes in the line profiles of {\it LCP} and {\it RCP} in the kernels ($K_2$, $K_4$ and, $K_5$) during the transients can be explained in terms of a combined change in DV, $B_0$ and line depth. 

As we are only interested in synthesis of the line profiles for a given change in different atmospheric parameters, we consider the inversion code ``MELANIE" which is based on Milne-Eddington (Skumanich \& Lites~\cite{Skumanich1987}) model. Using this code, the atmospheric parameters (magnetic field strength, Doppler velocity, Doppler width, line strength, damping constant, source function, source gradient and macro turbulence) were adjusted in such a way that simulated profiles of {\it LCP} and {\it RCP} derived from modeled Stokes {\it I} and {\it V} closely matches the observed profiles of {\it LCP} and {\it RCP} at the time of velocity and magnetic transients at each kernel.

\begin{figure*}[!t]
\begin{center}
\includegraphics[width=0.275\textwidth, angle = 90]{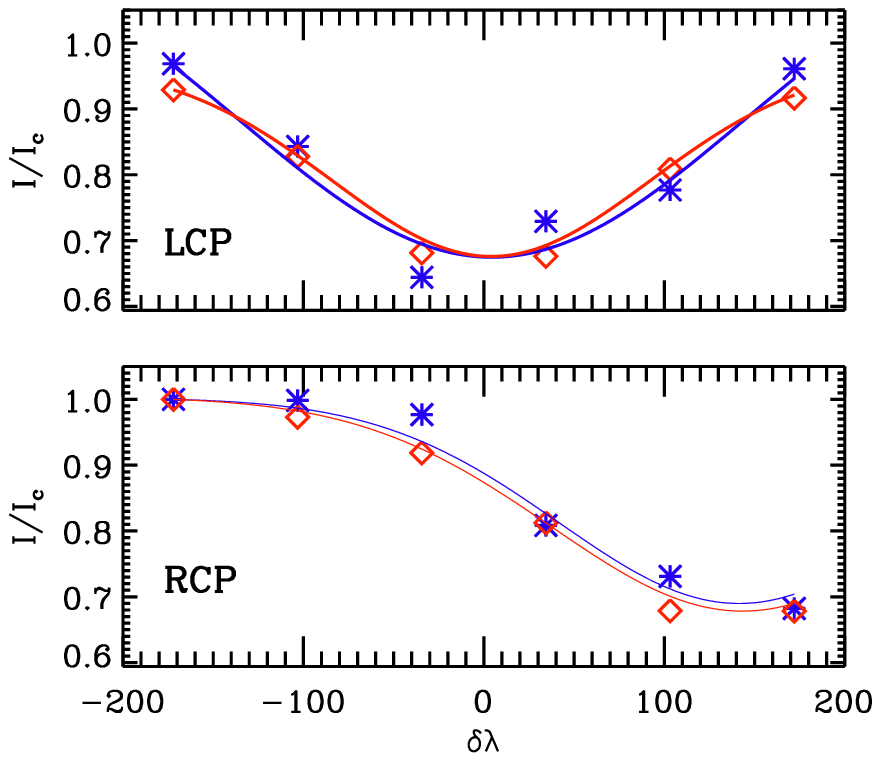}
\hspace{1mm}
\includegraphics[width=0.275\textwidth, angle = 90]{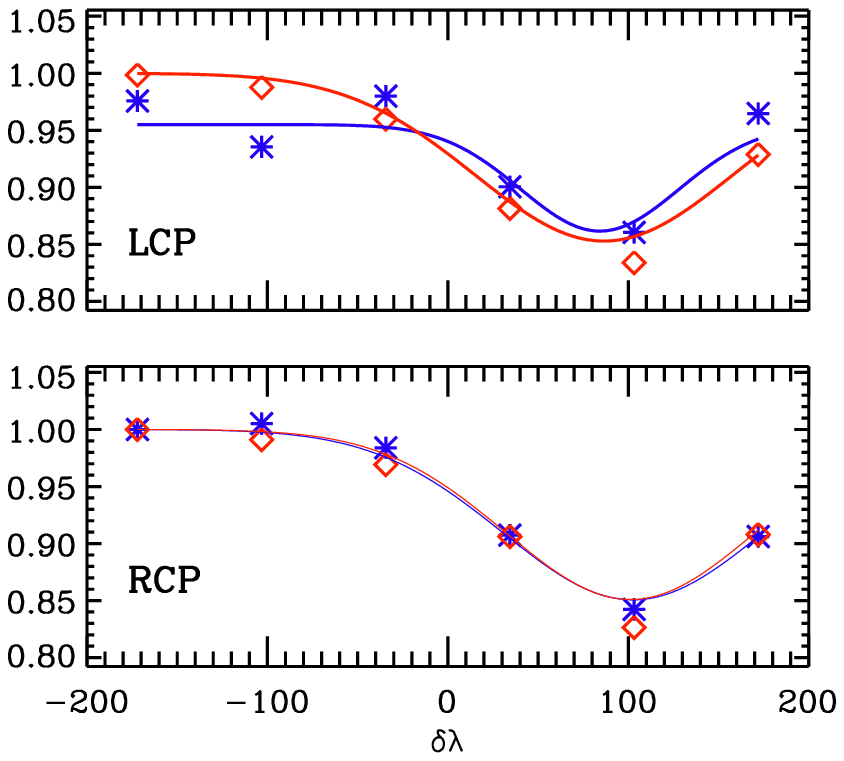}
\hspace{1mm}
\includegraphics[width=0.275\textwidth, angle = 90]{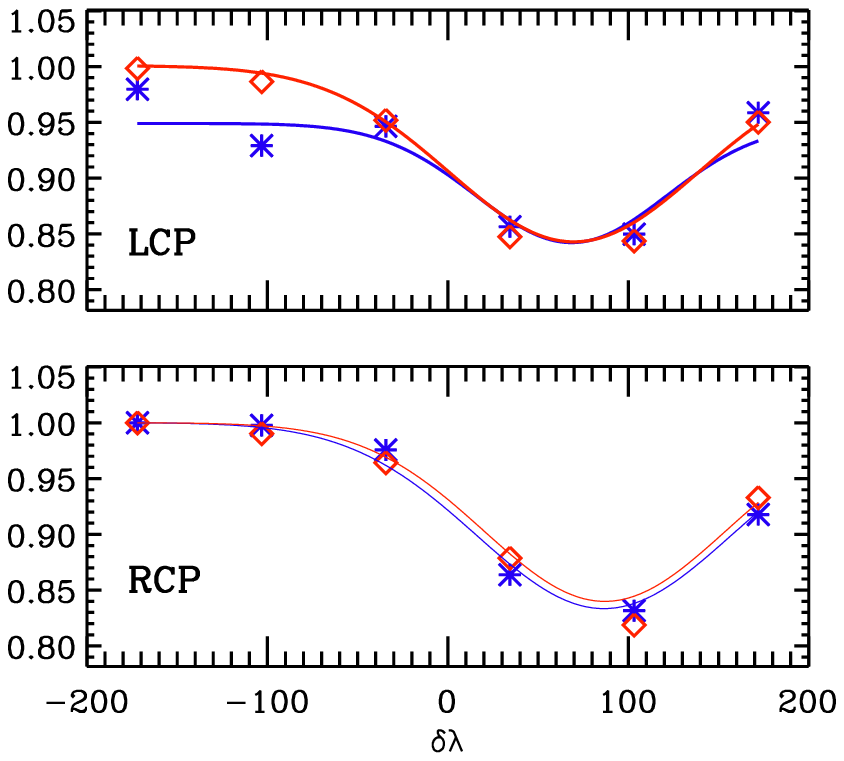}
\caption{Comparison of observed (blue) and synthesized (red) profiles of {\it LCP} and {\it RCP} for the kernel $K_2$ is shown in top and bottom panels, respectively, at the epochs: before the flare (left), at the time of velocity transient (middle) and at the time of magnetic transient (right). Symbols represent the normalized intensity at different wavelength positions and continuous line is for corresponding Gaussian fit.}
\label{K2BEF}
\end{center}
\end{figure*}
\begin{figure*}[!h]
\begin{center}
\vspace{-8mm}
\includegraphics[width=0.275\textwidth, angle = 90]{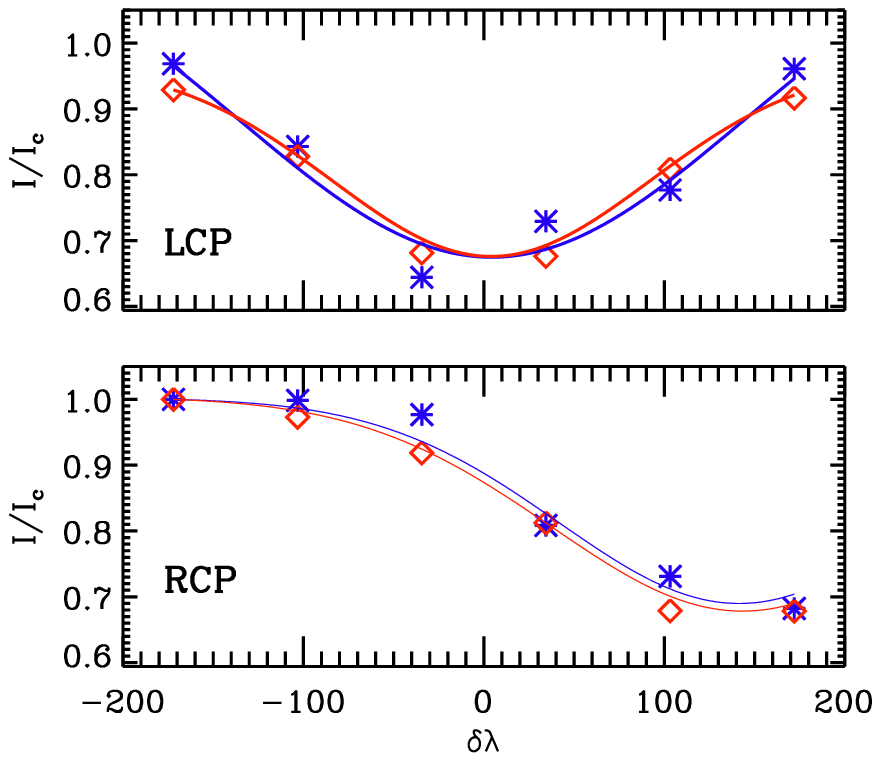}
\hspace{1mm}
\includegraphics[width=0.275\textwidth, angle = 90]{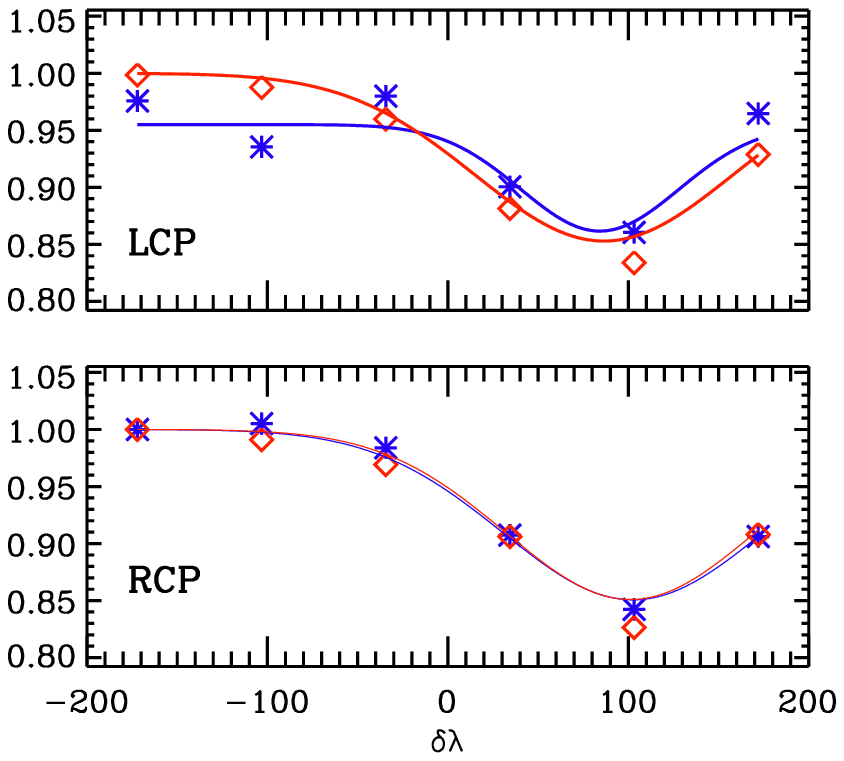}
\hspace{1mm}
\includegraphics[width=0.275\textwidth, angle = 90]{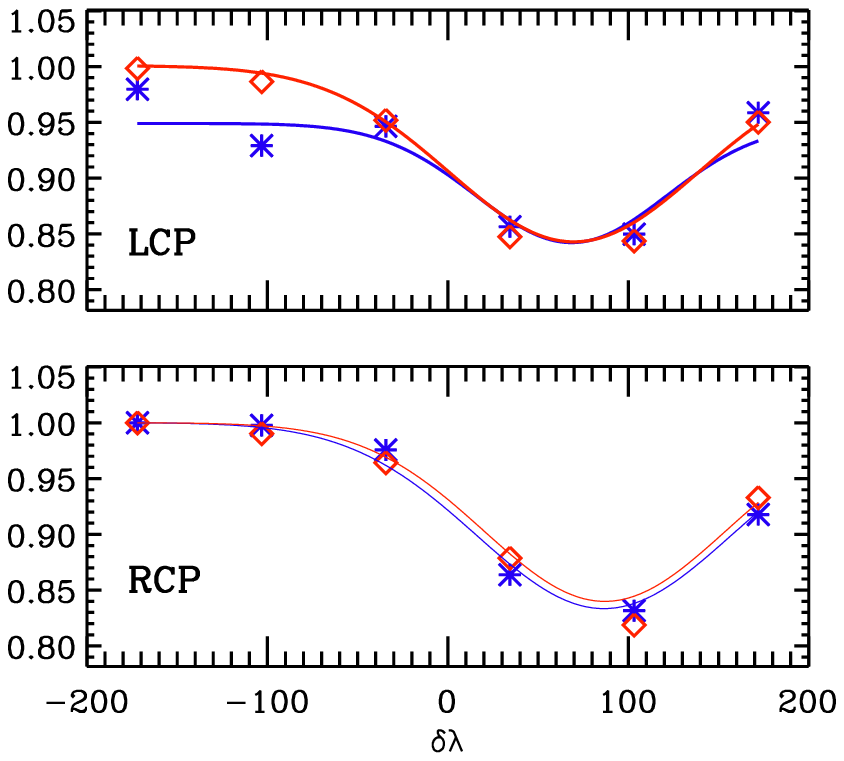}
\vspace{-3mm}
\caption{Same as figure~\ref{K2BEF} but for kernel $K_4$.}
\label{K4BEF}
\end{center}
\end{figure*}
\begin{figure*}[!h]
\begin{center}
\vspace{-5mm}
\includegraphics[width=0.275\textwidth, angle = 90]{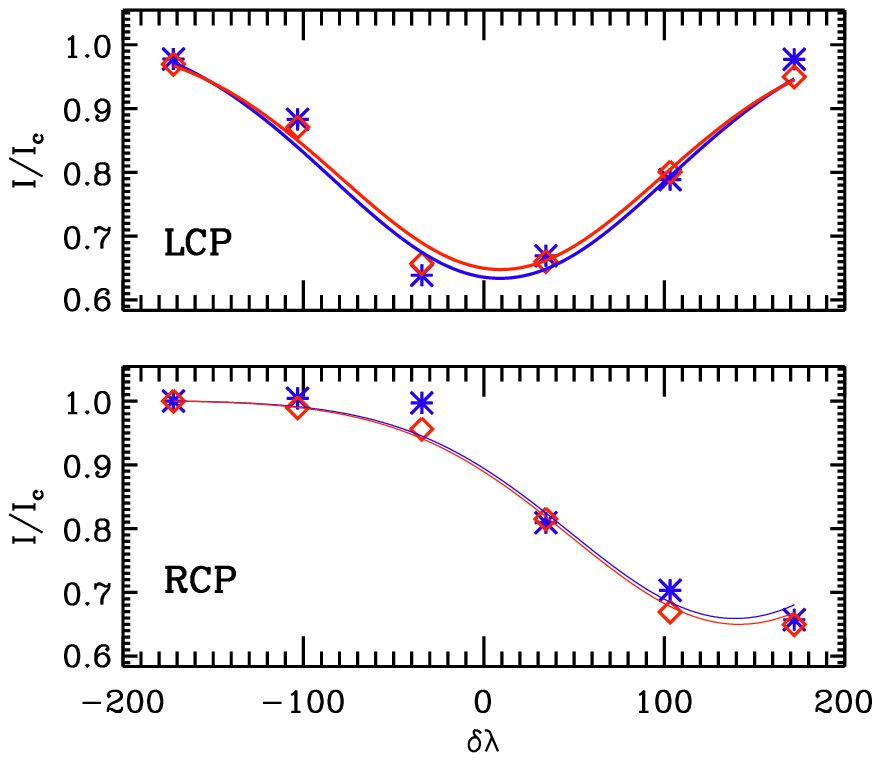}
\hspace{1mm}
\includegraphics[width=0.275\textwidth, angle = 90]{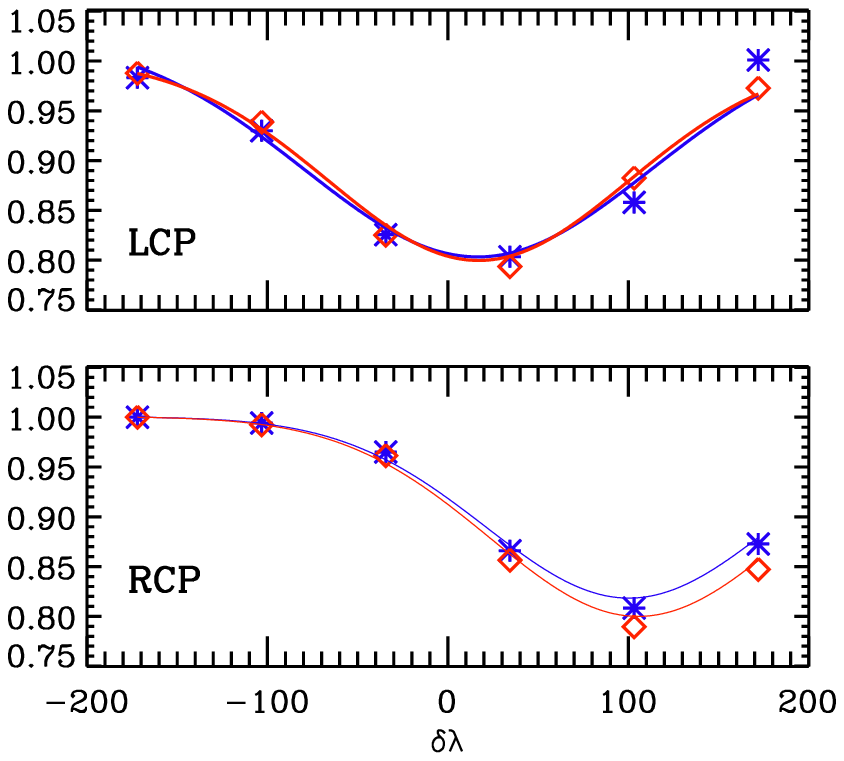}
\hspace{1mm}
\includegraphics[width=0.275\textwidth, angle = 90]{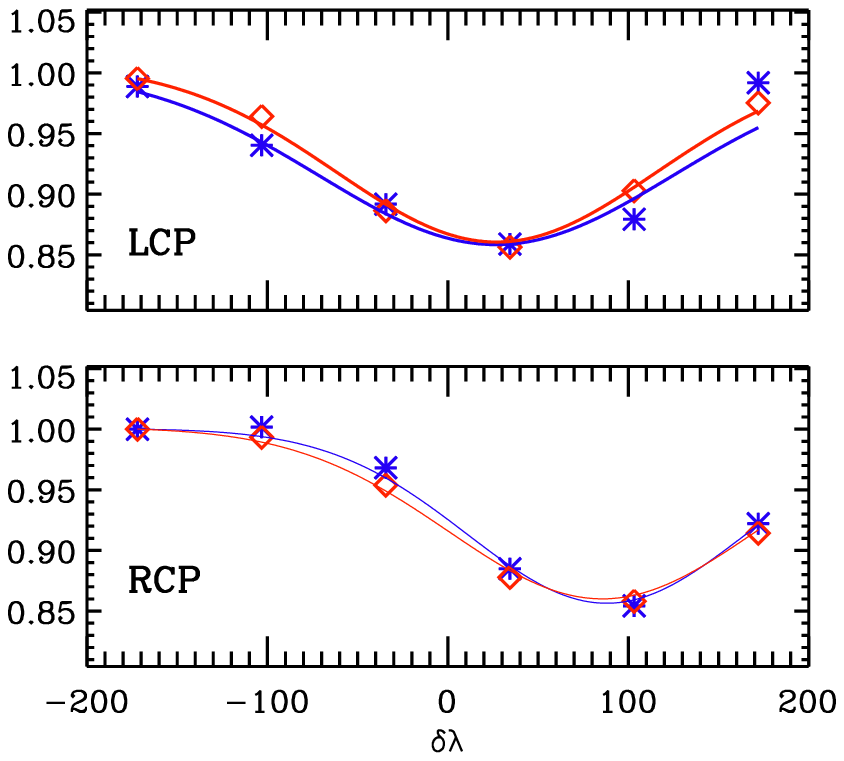}
\vspace{-3mm}
\caption{Same as figure~\ref{K2BEF} but for kernel $K_5$.}
\label{K5BEF}
\end{center}
\end{figure*}

In case of $K_2$, the dominant changes seen are the change in magnetic field strength of about 1400~G and change in velocity of about 1100~m/s in addition to change in line strength and source function. For $K_4$, a change in $B_0$ of about 1000~G and a change in $DV$ of about 500 m/s were the dominant changes in the atmospheric parameters in addition to minor changes in line strength, source function and inclination. Similarly, for $K_5$, the dominant changes seen are the change in $B_0$ of about 700~G and change in $DV$ of about 500~m/s in addition to change in line strength, source function and inclination. The synthesized line profiles along with observed ones for the kernels $K_2$, $K_4$ and $K_5$ at the time of velocity and magnetic transients are shown in Figures~\ref{K2BEF},~\ref{K4BEF}, and~\ref{K5BEF}, respectively. In each of these figures, the top row compares observed and synthesized profiles of {\it LCP } for the epochs: before the flare, at the time of velocity transient and at the time of magnetic transient, respectively, from left to right.  Similarly, the bottom row compares observed and synthesized profiles of {\it RCP}. It is observed that the synthesized profiles are in good agreement with the observed profiles of {\it LCP} and {\it RCP}. However, deviations are also seen in the {\it LCP} profile of $K_2$ at the time of velocity and magnetic transients and in the {\it RCP} profile of $K_4$ at the time of velocity transient.

We reconstructed profiles of Stokes {\it I} from the profiles of {\it LCP} and {\it RCP} to check for any emission or central reversal of the line profile during the transients (c.f.,~Figure~9). Here, we do not observe any emission or reversal in the line profiles for these kernels. In summary, forward modeling of the line profiles at the time of transients shows that a significant change in magnetic field strength and Doppler velocity relative to pre-flare condition are necessary in achieving the correlation between observed and synthesized profiles at all the kernels. 

\begin{figure*}[!h]
\begin{center}
\includegraphics[width=0.4\textwidth, angle = 90]{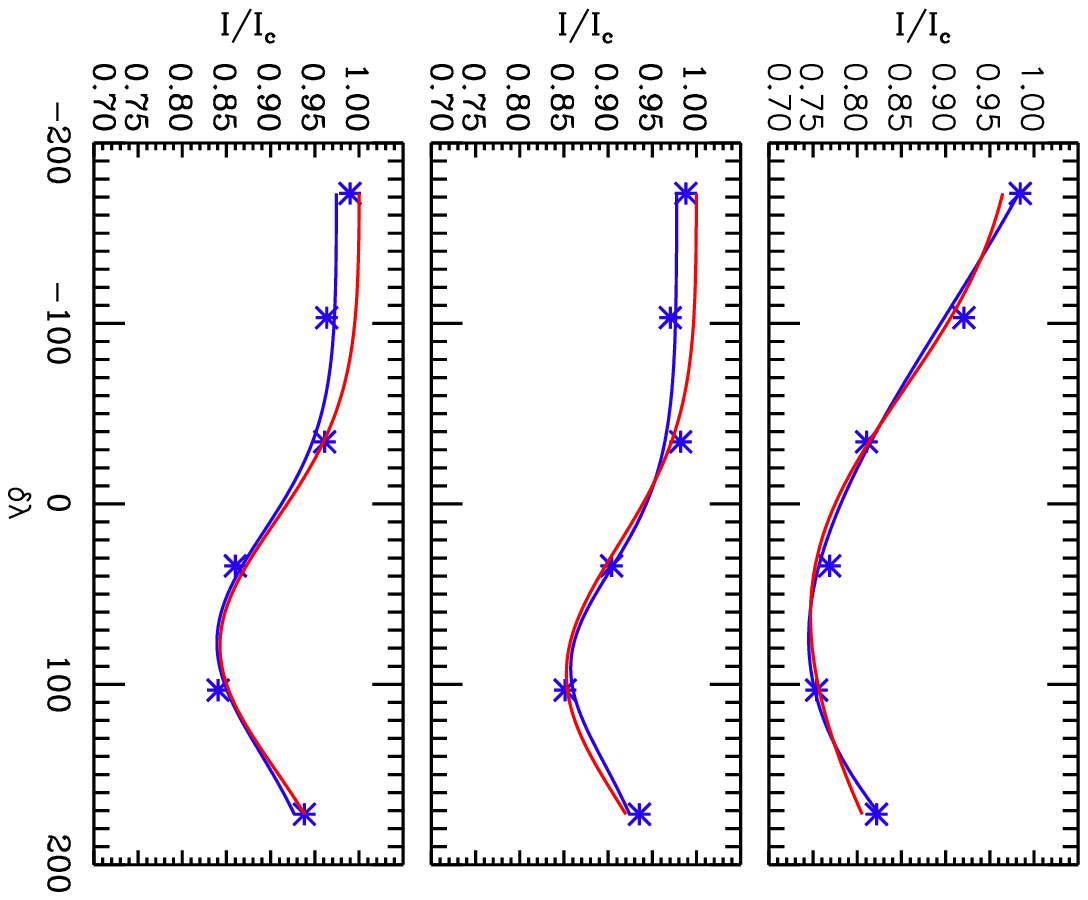}
\hspace{1mm}
\includegraphics[width=0.4\textwidth, angle = 90]{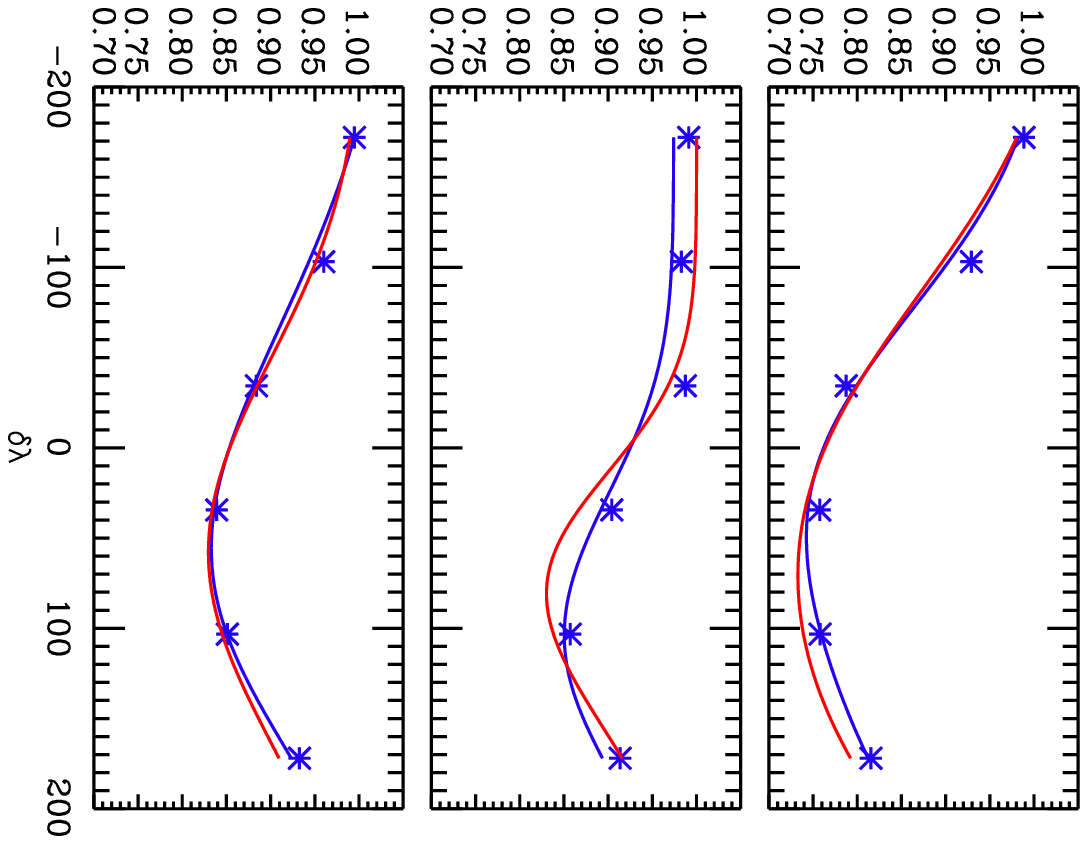}
\hspace{1mm}
\includegraphics[width=0.4\textwidth, angle = 90]{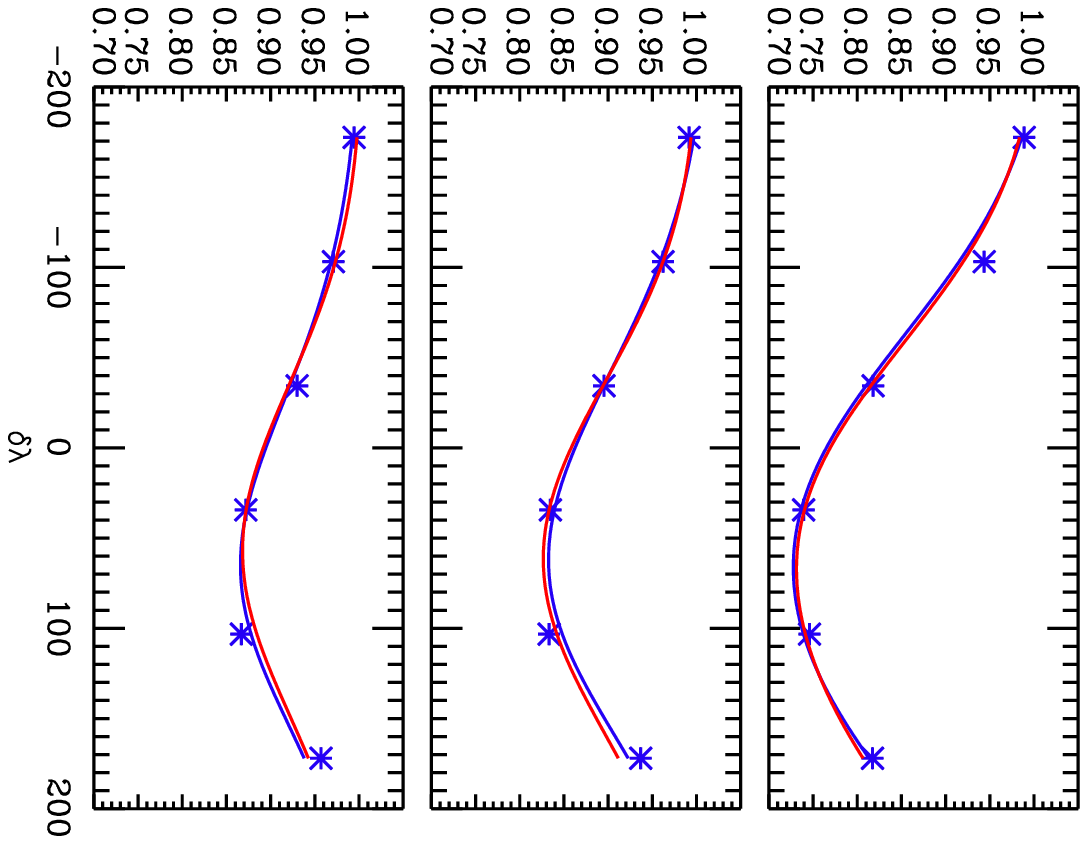}
\caption{Comparison of observed (blue) and synthesized (red) profiles of I (derived from profiles {\it LCP} and {\it RCP}) for the kernel $K_2$, $K_4$, and $K_5$ (left to right) at the epochs: before the flare (top), at the time of velocity transient (middle) and at the time of magnetic transient (bottom). Symbols represent the normalized intensity at different wavelength positions and continuous line is for corresponding Gaussian fit.}
\label{QaG}
\end{center}
\end{figure*}

\section{Discussion and Conclusions}

We have analyzed the response of the solar photospheric velocity flows and associated magnetic fields to the first major flare (of class X2.2) of the current solar cycle that occurred in the active region NOAA 11158 on 2011 February 15. This flare event was well observed by HMI instrument onboard {\it SDO} spacecraft. The HMI instrument provides high-resolution imaging spectroscopy of the active region, including continuum, Doppler, and magnetic maps of the solar disk in the photospheric Fe~I line centered at 6173~\AA~. Such features in the HMI data have enabled us to investigate the flare associated changes in velocity and magnetic signals observed in the active region NOAA 11158 on 2011 February 15. In this study, our chief findings are as follows:

\begin{enumerate}[i.]

\item The analysis of Dopplergrams of the active region spanning the flare event shows five kernels $K_1, K_2, K_3, K_4$, and $K_5$ in the active region which contain flare associated enhanced velocity signals. These kernels are located in and around the flare ribbons as seen in Ca~II~H images from SOT/FG onboard {\it Hinode}. It is also observed that all these kernels, except $K_1$, are spatially distributed within the HXR brightened location from RHESSI. Among these five kernels, some coincide with seismic sources shown by Kosovichev~(\cite{kosovichev2011}) and Zharkov~(\cite{Zharkov13}) and transients shown by Maurya et al.~(\cite{Maurya2012}). The kernels show enhanced downflows/upflows with a difference between minimum and maximum velocity being about 1 km/s. 

\item Identified kernels show velocity transients during the impulsive phase of the flare as seen in the {\it RHESSI} hard X-ray. A comparative study of the temporal evolution of hard X-ray light curve in the energy band 12-25~keV from {\it RHESSI} and velocity transients seen in the identified kernels $K_1, K_2, K_3, K_4$, and $K_5$ show that, except for $K_1$, in all other kernels the enhanced velocity signals appear within one minute after the peak time of the hard X-ray in the respective kernels. Thus, it can be assumed that these velocity transients are the photospheric responses of the energetic particles impinging on the solar photosphere during the flare as modeled by Zharkova \& Zharkov~(\cite{ZharZhar07}). However, the structured nature of the peak of hard X-ray light curve for $K_1$ makes it difficult to conclude whether the velocity impulse is generated as a result of particle impact. Kosovichev~(\cite{kosovichev2011}) proposes that for this kernel, the energy transport into the lower atmosphere could have been provided by the saturated heat flux. 

\item The study of magnetic field variations in the identified locations show that in most of these locations, the magnetic field after transients is different from the field before the transient. These observations similar to the earlier reports (Zhao et al.~\cite{Zhao09}, Wang et al.~\cite{Wang2009}, Wang et al.~\cite[and references therein]{Wang2011}) for several M-class and X-calss flares.

\item 
The study of the two circular polarization profiles ({\it LCP} and {\it RCP}) in the identified kernels shows that three out of the five locations show a change in line shape (depth and width) at the time of transients in comparison to that of the pre-flare condition. In order to understand the cause of these profile changes, we synthesized the line profiles using Milne-Eddington atmospheric model. We started with atmospheric parameters that represent pre-flare condition and showed that a considerable change in Doppler velocity, magnetic field and source function can lead to the observed changes in {\it LCP} and {\it RCP} profiles of the particular kernel at the time of both velocity and magnetic transients. We understand that the simultaneous changes in magnetic and velocity fields are necessary in addition to change in source function to cause such changes in the line profiles, particularly close to the points of inflection. 

However, it is commonly believed that the magnetic field measurements are greatly affected by line profile changes during the flares. The simulations by Qiu \& Gary~(\cite{QaG2003}) suggest that a magnetic reversal is mainly caused by line turning into emission from absorption and line broadening during the flares. Here, in our analysis we re-constructed profiles of Stokes {\it I} from the {\it LCP} and {\it RCP} measurements and we do not find any central reversal or emission at the time of transients (c.f., Figure 9). Additionally, the line profiles do not show any increase in line width. The location of the kernels in the periphery (c.f., Figure~2) of the HXR footpoints, with presumably less energy of impacted particle beam could be one reason for the absence of central reversal or emission in the observed Stokes {\it I} profiles. Therefore, we conclude that the changes in the line-of-sight observables seen in the kernels could well be real changes. 

In this regard, simultaneous measurements of {\it LCP} and {\it RCP} at a better cadence in the future can lead to a better understanding of such transients of velocity and magnetic fields. The best case scenario would be obtaining observations from a slit-based spectrograph with the flaring locations coinciding with the slit.

\end{enumerate}

\begin{acknowledgements}
This work utilizes data from the Helioseismic and magnetic Imager (HMI) onboard {\it Solar Dynamics Observatory} ({\it SDO}). This work also utilizes hard X-ray data from {\it RHESSI}, and Ca~II~H data from the Solar Optical Telescope (SOT) onboard {\it Hinode} mission. Savita Mathur acknowledges support from the NASA grant NNX12AE17G. R.~A.~Garc\'ia thanks the support of the GOLF CNES grant at the SAp/CEA-saclay. We are grateful to Dr. S. Couvidat for providing the spectral data from HMI used in our analysis and related discussions. We are thankful to the anonymous referee for useful comments and suggestions that improved the presentation of our manuscript. 
\end{acknowledgements}

\label{lastpage}

\end{document}